\documentclass[useAMS,usenatbib,usegraphicx]{mn2e}
\usepackage{amsmath,amsfonts,amssymb}

\def\Msun{\hbox{$\rm\, M_{\odot}$}}

\title [Evolution of the Galaxy Population]
{Confronting theoretical models with the observed evolution of the galaxy
population out to z=4}

\author[Henriques et al.]  
{Bruno Henriques$^{1}$\thanks{E-mail:bhenriques@mpa-garching.mpg.de},
Simon White$^{1}$,  Gerard Lemson$^{1}$, Peter Thomas$^{2}$, 
  \newauthor
  Qi Guo$^{3,4}$, Gabriel-Dominique Marleau$^{1,5}$, Roderik Overzier$^{1}$\\
  {}$^{1}$Max-Planck-Institut f\"ur Astrophysik, Karl-Schwarzschild-Str. 1, 85741 Garching b. M\"unchen, Germany\\
  {}$^{2}$Astronomy Centre, University of Sussex, Falmer, Brighton BN1 9QH, United Kingdom\\
  {}$^{3}$Partner Group of the Max-Planck-Institut f\"ur Astrophysik, National Astronomical Observatories, Chinese Academy of Sciences, \\
             ~Beijing, 100012, China\\
  {}$^{4}$Department of Physics, Institute for Computational Cosmology, University of Durham, South Road, Durham DH1 3LE\\
  {}$^{5}$Department of Physics, McGill University, 3600 Rue University, Montr\'eal, QC H3A 2T8, Canada\\
}
 
\begin{document}

\date{Submitted to MNRAS}

\pagerange{\pageref{firstpage}--\pageref{lastpage}} \pubyear{2011}

\maketitle

\label{firstpage}

\begin{abstract}
  We construct lightcones for the semi-analytic galaxy formation
  simulation of \citet{Guo2011} and make mock catalogues for
  comparison with deep high-redshift surveys. Photometric properties
  are calculated with two different stellar population synthesis codes
  \citep{Bruzual2003, Maraston2005} in order to study sensitivity to
  this aspect of the modelling. The catalogues are publicly available
  and include photometry for a large number of observed bands from
  $4000\AA$ to $6\mu{\rm m}$, as well as rest-frame photometry and
  other intrinsic properties of the galaxies (e.g positions, peculiar
  velocities, stellar masses, sizes, morphologies, gas fractions, star
  formation rates, metallicities, halo properties).  \citet{Guo2011}
  tuned their model to fit the low-redshift galaxy population but
  noted that at $z\geq 1$ it overpredicts the abundance of galaxies
  below the ``knee'' of the stellar mass function. Here we extend the
  comparison to deep galaxy counts in the $B$, $i$, $J$, $K$ and IRAC
  $3.6\rm{\mu m}$, $4.5\rm{\mu m}$ and $5.8\rm{\mu m}$ bands, to the
  redshift distributions of $K$ and $5.8\rm{\mu m}$ selected galaxies,
  the evolution of rest-frame luminosity functions in the $B$ and $K$
  bands and the evolution of rest-frame optical versus near-infrared
  colours. The $B$, $i$ and $J$ counts are well reproduced, but at
  longer wavelengths the overabundant high-redshift galaxies produce
  excess faint counts. At bright magnitudes, counts in the IRAC bands
  are underpredicted, reflecting overly low stellar metallicities and
  the neglect of PAH emission. The predicted redshift distributions
  for $K$ and $5.8\rm{\mu m}$ selected samples highlight the effect of
  emission from thermally pulsing AGB stars.  The full treatment of
  \citet{Maraston2005} predicts three times as many $z\sim 2$ galaxies
  in faint $5.8\rm{\mu m}$ selected samples as the model of
  \citet{Bruzual2003}, whereas the two models give similar predictions
  for $K$-band selected samples.  Although luminosity functions are
  adequately reproduced out to $z\sim 3$ in rest-frame $B$, the same
  is true at rest-frame $K$ only if TP-AGB emission is included, and
  then only at high luminosity.  Fainter than L$_\star$ the two
  synthesis models agree but overpredict the number of galaxies,
  another reflection of the overabundance of $\sim10^{10}M_\odot$
  model galaxies at $z\geq 1$. The model predicts that red, passive
  galaxies should already be in place at $z=2$ as required by
  observations.

\end{abstract}


\section{Introduction}
\label{sec:intro}

Semi-analytic models of galaxy formation aim to predict the evolution
of population properties such as the distributions of stellar mass,
luminosity, star formation rate, size, rotation velocity, morphology,
gas content and metallicity, as well as the scaling relations linking
these properties. They follow astrophysical processes affecting the
baryonic components using a series of analytic, physically based
models which are embedded either in an analytic representation
\citep{White1991,Kauffmann1993, Cole1994,Somerville1999} or in a
direct numerical simulation \citep{Kauffmann1999, Springel2001,
  Springel2005} of the evolution of the underlying dark matter
distribution. Uncertain efficiencies and scalings of these
astrophysical processes are represented by adjustable
parameters. These may be set {\it a priori} through a detailed
calculation or simulation of specific processes, or they may be
determined observationally by matching suitably chosen data
(e.g. \citealt{Croton2006}; \citealt{Bower2006}; \citealt{Menci2006};
\citealt{Cattaneo2006}; \citealt{DeLucia2007}; \citealt{Monaco2007};
\citealt{Somerville2008}; \citealt{Guo2011}). The extremely broad
range of relevant data and the considerable freedom in specifying
appropriate recipes complicate the systematic comparison of
semi-analytic models with data. New and robust statistical tools have
recently been developed to facilitate quantitative comparisons
\citep{Kampakoglou2008, Henriques2009, Henriques2010, Bower2010,
  Lu2010}.

Such comparisons are sensitive to stellar population synthesis models
which are required both to derive intrinsic galaxy properties such as
mass, age and star formation rate from observational data, and to
calculate luminosities, colours and spectra for model
galaxies. Erroneous conversions between physical and observable
properties lead to incorrect conclusions about galaxy formation
physics, so it is important to check the implications of adopting
differing stellar populations synthesis models
(e.g. \citealt{Buzzoni1989}; \citealt{Worthey1994};
\citealt{Vazdekis1996}; \citealt{Fioc1997}; \citealt{Leitherer1999};
\citealt{Bruzual2003}; \citealt{Thomas2003}; \citealt{Maraston2005};
\citealt{Conroy2009}).

For example, the impact of including models for thermally-pulsating
asymptotic giant branch (TP-AGB) stars has been studied in some detail
in recent years. The contribution from these stars can significantly
enhance the near-infrared emission of galaxies with Gyr-old
populations (e.g.  \citealt{Maraston1998}; \citealt{Maraston2005};
\citealt{Wel2006}, \citealt{Marigo2007}; \citealt{Charlot2007};
\citealt{Conroy2009}). The data of \citet{Conroy2009},
\citet{Marchesini2009, Marchesini2010}, \citet{Zibetti2009} and
\citet{Santini2011} show that inclusion of this additional emission
can reduce the masses inferred from K-band light by as much as 0.6
dex. The semi-analytic models of \citet{Tonini2009, Tonini2010},
\citet{Fontanot2010} and \citet{Henriques2011} suggest that a
substantial contribution from TP-AGB stars, as predicted by the model
of \citet{Maraston2005} may explain the large number of extremely red
objects found at $z\sim 2$. Other examples of how uncertainties in
stellar evolution modelling affect physical inferences from data are
given by \cite{Conroy2010}.  Here we compare predictions of the
\citet{Guo2011} semi-analytic model for two different population
synthesis models, the one originally used by these authors (from
\citealt{Bruzual2003}) and that of \citet{Maraston2005}.

Even neglecting uncertainties from stellar population modelling,
mass-to-light ratios and other physical properties are often poorly
constrained by available data. Estimates rely on fitting theoretical
models to observed photometry and spectral energy distributions (SEDs)
and approximately equivalent fits can often be obtained for broad
ranges of assumed star formation history, chemical enrichment history
and obscuration by dust. Additional uncertainties arise from possible
variations in the initial mass function (IMF) with which stars form,
and from possible spectral contributions from an active galactic
nucleus (AGN).  It has been argued in the past that a single optical
colour (e..g. $g-i$) is, in practice, sufficient to derive
light-to-mass ratios accurate to 0.1 dex for most galaxies
\citep{Bell2003, Gallazzi2009}.  However, \citet{Zibetti2009} showed
that this is only true for relatively weak obscuration. For heavily
obscured young populations, resolved photometry is needed to achieve
an accuracy better than 0.2~dex and even that requires an additional
near-infrared colour (e.g. one may combine $g-i$ and $i-H$).

Galaxy formation models directly predict star formation and enrichment
histories, so in the absence of obscuration a well defined SED can be
predicted for each galaxy as a superposition of simple stellar populations
(SSPs), each made up of coeval stars of a single metallicity.  The ``observed''
photometry is then easily obtained by redshifting the SED and integrating over
the appropriate photometric filter functions. In practice, however, the
conversion to observables is heavily influenced by dust and is sensitive to
the details of its distribution within a galaxy (e.g. \citealt{Granato2000,
  Cole2000}). This significantly limits the precision with which
observables can be predicted from galaxy formation models. Current
semi-analytic models often attempt to handle these uncertainties by using
observational data to constrain the dust model \citep[e.g.][]{Granato2000,
Cole2000,  Kitzbichler2007, Guo2009}.

%In a comprehensive study of the complications that arise when
%converting theoretical predictions into observables,
%\citet{Conroy2010} describe the uncertainties in stellar evolution
%theory and dust corrections.  These uncertainties lead the authors
%into preferring the conversion from light to mass. However, and has
%shown by the same authors in the first paper of a series
%\citep{Conroy2009}, stellar evolution theory has an even larger impact
%in the derivation of properties from observations. Particularly at
%high redshift when only limited photometry is available. Moreover, as
%the authors pointed out, for current versions of semi-analytic models
%where the total amount of metals and gas content is predicted
%($\tau_2$ in their notation), the uncertainties in dust extinction
%only impact ultra-violet colors.

Because of such difficulties it seems wise to compare theory and
observation for a broad range of properties, at different redshifts,
and at different ``conversion levels''. The latter is particularly
crucial at high-redshift, where very limited data are available and
the relations between mass, light and star-formation rate are very
uncertain. There is no preferred ``comparison frame'' and conclusions
are convincing only if a consistent picture emerges which matches
smoothly onto the lower redshift galaxy populations. \citet{Guo2011}
compare their model extensively to low-redshift galaxies but only
present limited predictions at high redshift. Specifically, they
compare to published estimates of the evolution of the stellar mass
function of galaxies out to $z\sim 4$, finding significant
discrepancies for stellar masses below $5\times 10^{10}\Msun$.
%The authors find a good agreement between
%model and data at the high mass end and an excessive number of dwarfs
%at high redshift. We note that the agreement on the high mass end is
%only achieved by assuming additional systematic errors in observations
%not considered in the data. We will show that the model can indeed fit
%the photometry used to derive the mass functions and that the
%discrepancy is likely due to erroneous conversions from light to
%mass. 

In this paper we extend this comparison considerably, analyzing the
photometric properties of galaxies from high redshift to the present
day, and comparing with observations at a variety of levels from
number counts as a function of apparent magnitude, through redshift
distributions of magnitude limited samples, to rest-frame luminosity
functions as a function of redshift. In particular, we study
predictions for the near-infrared bands for which data have recently
become available from the Spitzer satellite. To facilitate this work
we build lightcones using the MoMaF software \citep{Blaizot2005}. We
are then able to reproduce the observational selection criteria of
modern surveys such as the UKIRT Infrared Deep Sky Survey (UKIDSS)
\citep{Lawrence2007}, the VIMOS-VLT Deep Survey (VVDS)
\citep{LeFevre2005}, the Deep Evolutionary Exploratory Probe 2 Galaxy
Redshift Survey (DEEP2) \citep{Davis2003} or the Cosmic Evolution
Survey (COSMOS) \citep{Scoville2007}. 

Similar studies were performed by \citet{Kitzbichler2007},
\citet{Guo2009} and \citet{Torre2011} for earlier versions of the
Munich semi-analytic model.  The results here are based on the model
of \citet{Guo2011} which is implemented simultaneously on the
Millennium and Millennium-II Simulations and was retuned to fit a
broad range of ``high precision'' data on the low-redshift galaxy
population, primarily from the Sloan Digital Sky Survey (SDSS). We
also expand the photometric coverage from the ultraviolet to IRAC
bands and test the dependence on stellar population synthesis
modelling over this wavelength range. In a recent paper,
\citet{Somerville2011} compared a different semi-analytic model with
photometry extending to even longer wavelengths (the far
infrared). This required modelling the re-emission of starlight by
heated dust, as also considered by \citet{Granato2000},
\citet{Cole2000}, \citet{Baugh2005}, \citet{Lacey2010}. Here we avoid
this complication and compare only to directly observed starlight.

This paper is organized as follows. In Section \ref{sec:models} we
summarize the characteristics of the semi-analytic model we use and we
describe how we construct lightcones for it. Section \ref{sec:results}
then presents results for number counts and redshift distributions as
a function of apparent magnitude, and for rest-frame $B$- and $K$-band
luminosity functions as a function of redshift. In Section
\ref{sec:conclusions} we present our conclusions.

\section{The Semi-Analytic Model}
\label{sec:models}

Modern semi-analytic models are built on merger trees from
high-resolution dark matter simulations. These provide a description of
the evolution of the mass and number density of dark matter halos and
the subhalos within them, as well as of their spatial and kinematic
distributions. The evolution of the baryonic components hosted by
these (sub)halos is then followed using a set of simplified formulae
describing each of the relevant astrophysical processes. The latest
version of the Munich model \citep{Guo2011} is implemented on two very
large dark matter simulations, the Millennium Simulation
\citep[MS;][]{Springel2005} and the Millennium-II Simulation
\citep[MS-II;][]{Boylan2009}. The MS follows the evolution of
structure within a cube of side $500h^{-1}\rm{Mpc}$ (comoving) and its
merger trees are complete for subhalos above a mass resolution limit
of $1.7\times10^{10}h^{-1}\rm{M}_{\sun}$. The MS-II follows a cube of
side $100h^{-1}\rm{Mpc}$ but with 125 times better mass resolution
(subhalo masses greater than $1.4\times10^{8}h^{-1}\rm{M}_{\sun}$).
Both adopt the same WMAP1-based cosmology \citep{Spergel2003} with
parameters $h=0.73, \Omega_m=0.25, \Omega_\Lambda=0.75, n=1$ and
$\sigma_8=0.9$. These are outside the region preferred by more recent
analyses (in particular, $\sigma_8$ appears too high) but this is of
no consequence for the issues we study in this paper. For consistency,
we will use this cosmology whenever it is necessary to derive the
physical properties of galaxies from observed fluxes and
redshifts. The distributions of physical properties converge in the
two simulations for galaxies with
$10^{9.5}\rm{M}_{\sun}<\rm{M}_{\star}<10^{11.5}\rm{M}_{\sun}$. In this
study we focus only on results from the MS, since its resolution limit
is well below the stellar masses covered by the datasets with which we
compare.

\subsection{The model of Guo et al. 2011}
\label{sec:guo2011}

For a full description of the semi-analytic model used in this work we refer
the reader to \citet{Guo2011}. Here we briefly describe changes from earlier
versions of the Munich semi-analytic model that significantly affect our
results.

Following \citet{Kitzbichler2007} and \citet{Guo2009}, the model of
\citet{Guo2011} includes a redshift-dependent model for internal
extinction which assumes that the dust-to-gas ratio increases with
metallicity but decreases with redshift. The effective optical depth
is given by:

\begin{equation} \label{eq:msamextinctionism}
\textstyle {\tau_{\lambda}=\left(\frac{A_{\lambda}}{A_{\rm{v}}}\right)_{Z_{\odot}}(1+z)^{-0.4}
\left(\frac{Z_{\rm{gas}}}{Z_{\odot}}\right)^s\left(\frac{\langle N_H\rangle}{2.1\times10^{21}{\rm{atms}} \,{\rm{cm}}^{-2}}\right)},
\end{equation}
where $\langle N_H\rangle$ represents the mean column density of
hydrogen, $(A_{\lambda}/A_{\rm{v}})_{Z_{\odot}}$ is the extinction curve for
the solar metallicity taken from \citet{Mathis1983} and $s=1.35$ for
$\lambda<2000$~\AA $\:$ and $s=1.6$ for $\lambda>2000$ \AA.

When they implemented the \citet{DeLucia2007} version of the Munich
model on the high-resolution Millennium-II Simulation, \citet{Guo2011}
found it to overproduce dwarf galaxies. The authors therefore
increased the efficiency of supernova feedback by introducing a direct
dependence of the amount of gas reheated and ejected on the virial
mass of the host halo. However, although the resulting model fits the
stellar mass function of galaxies well at low redshift, it still
produces more low-mass galaxies than are observed at $z>1$. This
deficiency is reflected in our results below.

Finally, \citet{Guo2011} introduced a more realistic treatment of
satellite galaxy evolution and of mergers. The hot gas content of
satellite galaxies is gradually stripped instead of being
instantaneously removed at infall, as suggested by the simulations of
\citet{McCarthy2008}. This allows satellites to continue forming stars
for a longer period and reduces the excessively rapid reddening of
these objects. In addition, satellites of satellites remain connected
to their parent galaxies and can merge with them, rather than being
automatically reassigned to the central galaxy of the group or
cluster. The model also includes a treatment of the tidal disruption
of satellite galaxies.\footnote{See \citet{Henriques2010} for an
  alternative extension of the Munich semi-analytic model modifying
  supernova feedback and including tidal disruption of satellites.}
  
\subsection{Lightcone Construction}
\label{sec:lightcones}
  
At high redshift, the observed fluxes at a limited number of
wavelengths are often the only data available for a galaxy, so that
its redshift must be inferred through comparison of the observed
colours to model templates. Even rest-frame magnitudes, colours and
luminosities can then be subject to substantial uncertainties, and the
conversion to intrinsic properties such as masses and star formation
rates is problematic. Results not only depend on the accuracy of the
photometric redshift, but are also (almost) degenerate with respect to
the star formation history, metallicity and dust content of the
galaxy. These quantities are direct predictions of a semi-analytic
model, so that the conversion from intrinsic to observed properties
is, in principle, well defined, given a stellar population synthesis
model, an assumed IMF and a specific model for intrinsic
obscuration. It is thus often convenient to consider conversion
uncertainties as part of the model and to compare theoretical
predictions directly with observables. To do this, we construct
lightcones which allow the models to be ``observed'' in a way that
mimics real surveys as directly as possible.  We use two different
population synthesis models to predict observables in order to assess
the impact of the differing mass-to-light conversions they imply. We
compute observed- and rest-frame fluxes from the ultraviolet to the
near infrared so that our theoretical datasets resemble those of
modern observational surveys not only in volume, but also in
wavelength coverage.

Our lightcones are built using the Mock Map Facility (MoMaF) developed by
\citet{Blaizot2005}. We refer the reader to the original paper for a full
description of the method. Here we briefly summarize the complications that
arise when building lightcones from simulations of limited size and
resolution. 

The Millennium Simulation has side of $500h^{-1}$ Mpc (comoving).
This is considerably smaller than, for example, the comoving distance
to a galaxy observed at $z\sim 2$. Periodic replication of the
simulation can lead to multiple appearances of the same object within
the lightcone, although typically at different redshifts and so with
different properties and at offset positions (due to large-scale
motions). \citet{Blaizot2005} suggested applying a series of
transformations (rotations, translations and inversions) when tiling
space with periodic replications. This does not, of course, prevent
multiple appearances of a given object within the lightcone, but these
duplicates are then viewed from different directions and no longer
fall on a (nearly) regular lattice.

Unfortunately, this technique also introduces discontinuities in
large-scale structure at the boundaries between replications,
affecting clustering statistics in a way which is at least as
difficult to model as that of the original periodicity.
\citet{Kitzbichler2007} showed that for lightcones of relatively small
solid angle, the central line-of-sight can be chosen to pass through
the lattice of periodic replications in such a direction that multiple
images of the same object are minimised or eliminated altogether. The
latter is not possible if the comoving volume of the lightcone exceeds
that of the simulation, but this technique can still be used to ensure
that multiple appearances occur as far apart as possible both on the
sky and in redshift. We therefore use the method of
\citet{Kitzbichler2007} in this paper. Space is filled with periodic
replications of the simulation, a position is chosen for the observer,
and the central line-of-sight of the survey field is given a
previously chosen orientation.  Galaxies whose positions intercept the
lightcone are selected and their comoving distance is converted into a
redshift.

As explained in \citet{Kitzbichler2007}, the time between stored
snapshots for the Millennium Simulation varies between 100 and 380
Myr. This means that the intrinsic properties of galaxies are not
generally available at the time corresponding to their comoving
distance. Rather they must be taken from the stored snapshot which is
closest to their light-cone position. Hence, galaxies with redshift
$(z_i+z_{i-1})/2<z<(z_i+z_{i+1})/2$ are assigned the physical
properties stored at $z_i$.  The resulting discontinuity in galaxy
population properties, at the boundaries between snapshots, could be
reduced by interpolating, but this works poorly for positions and
velocities since the output separation is comparable to orbital times
within groups and clusters. Moreover, it is not straightforward for
other galaxy properties either since these change discontinuously on
timescales shorter than the output spacing, for example through
mergers and starbursts. We thus follow \citet{Kitzbichler2007} and do
not attempt any interpolation. The semi-analytic calculations
  are perfomed on these intermediate time-steps that vary between 5
  and 15 Myr. This means, for example, that a burst of star formation
  will have this duration and can happen anywhere between (or at)
  output snapshots, with the corresponding increase in flux being
  reflected in galaxy properties at the snapshot.

%Note that we do check all galaxies near the
 % boundaries between snapshots to ensure that peculiar motions do not
 % cause them to appear in the lightcone twice or to be omitted
 % altogether. Each friends-of-friends group of galaxies appears once
 % and only once in the lightcone with physical properties
 % corresponding to the snapshot for which the group centre is closest
 % in comoving distance to that of the snapshot.

The {\it apparent} luminosities and colours of galaxies depend
strongly on their redshifts through the conversion between rest- and
observed-frame photometric bands and through the inverse square
dependence of apparent luminosity on distance.  The final redshift of
the galaxy in the lightcone is not available at the time
observed-frame luminosities are computed in the semi-analytic
model. However, there will be two extreme redshifts that bracket
it. We compute apparent observed-frame luminosities (for fixed
intrinsic properties) using these upper and lower limits, and once the
galaxy is placed in the lightcone, we interpolate to obtain final
observed-frame quantities.

For this paper we construct lightcones for square areas of $1.4\times
1.4$ deg$^2$ out to high redshift with no faint magnitude cut. They
are however limited by the mass resolution of the dark matter
simulation ($1.7\times10^{10}h^{-1}\rm{M}_{\sun}$ in halo mass)
corresponding to stellar masses of $\sim10^{9.5}\rm{M}_{\sun}$ at
z=0. While this does not matter for the questions we study in this
paper, it should be borne in mind if the lightcones are used for other
purposes.\footnote{The lightcones are publicly available at
  http://www.mpa-garching.mpg.de/millennium.}

\subsection{Stellar Populations and  Photometry}
\label{sec:ssp}
  
\begin{figure}
\centering
\includegraphics[width=8.4cm]{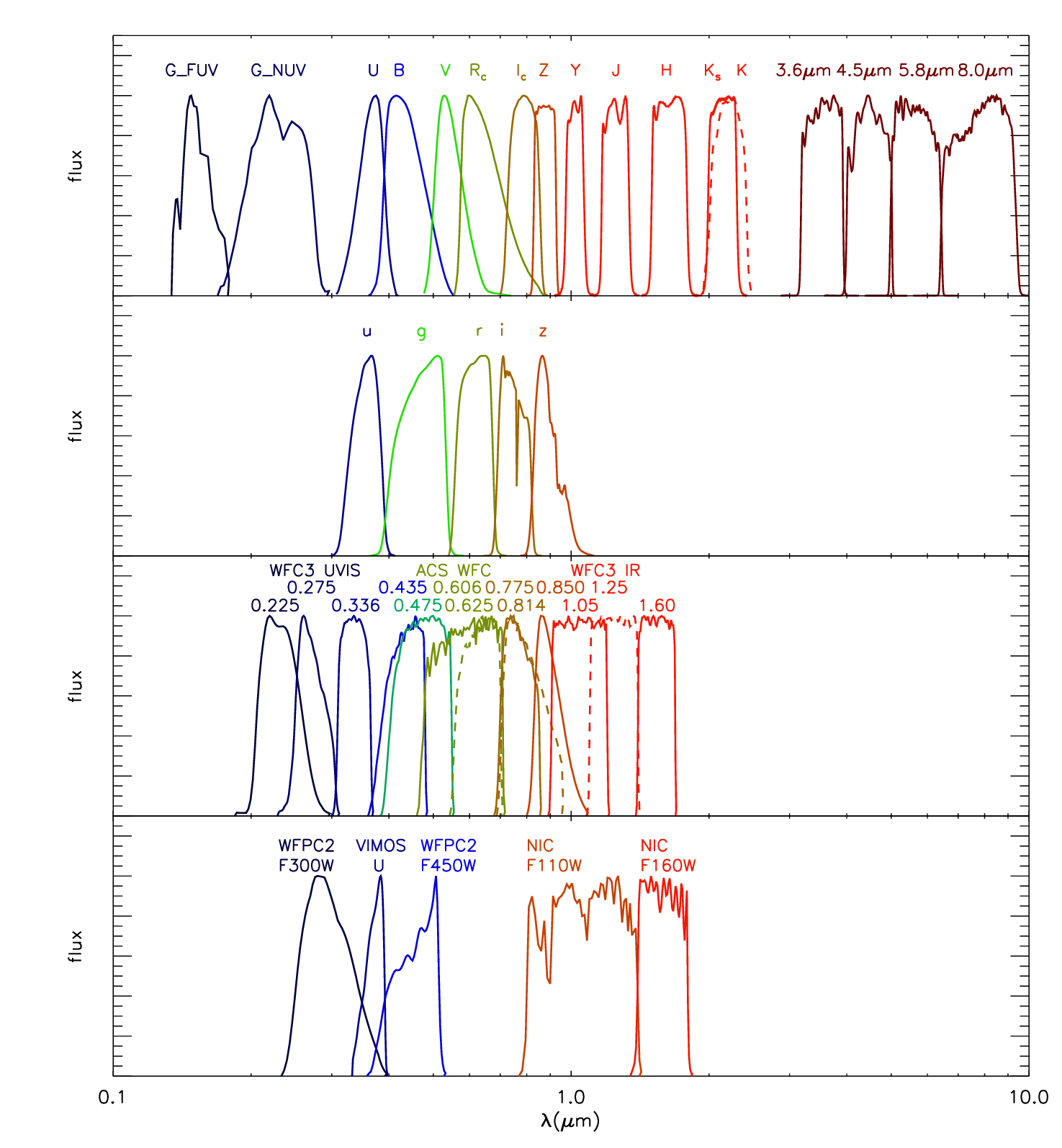}
\caption{The response functions of the filters for which fluxes are computed
  for the lightcones produced in this paper. These extend from the far
  ultra-violet to the IRAC bands and include: in the top panel, GALEX FUV and
  NUV, Johnson $UBVR_cI_cJHK_sK$ and IRAC bands; in the second panel, SDSS
  $ugriz$; in the third panel, HST WFC3 UV and IR and ACS WFC; and in the
  bottom panel, HST WFPC2, VIMOS U and NICMOS.}
\label{fig:filters}
\end{figure}

\subsubsection{Stellar Population Synthesis Models}

Semi-analytic models predict intrinsic properties of galaxies, such as
stellar mass, star formation history, gas and dust content and
metallicity. In order to convert these into observed spectral energy
distributions (SED) or broad-band photometry, evolutionary population
synthesis and dust models are required. The former predict the
evolution of the light associated with a single short burst of star
formation of given metallicity and with an assumed Initial Mass
Function (IMF), a so-called Simple Stellar Population (SSP). The
intrinsic stellar emission from a model galaxy is then represented as
a superposition of SSPs weighted according to its star formation
history. This emission must be processed through a dust model in order
to predict the observable stellar emission. Uncertainties in the
conversion between mass and light can jeopardise any comparison
between theory and observations. There are still significant
differences between published evolutionary population synthesis models
and these should be considered as part of the systematic uncertainties
when comparing semi-analytic model predictions to data. Throughout
this paper we present results from two distinct stellar population
synthesis codes: one that has been traditionally used in the Munich
model \citep{Bruzual2003} and the \citet{Maraston2005} model
implemented in the semi-analytic code by \citet{Henriques2011}.  In
both cases we adopt the same \citet{Chabrier2003} IMF and a similar
metallicity grid. We hope that the differences we find will give some
indication of the impact of mass-to-light conversion uncertainties on
galaxy formation modelling.

\subsubsection{Photometry}

In order to increase the predictive power of the model and allow it to
be tested against a wider range of observations, we also expand the
number of photometric bands for which fluxes are computed, covering
all wavelengths dominated by direct emission from stars, from the UV
to the near-infrared IRAC bands. In Fig. \ref{fig:filters} we plot the
relevant filter transmission curves. In the top panel we show the
GALEX FUV and NUV, the Johnson $U, B, V, R_c, I_c, Z, Y, J, H, K_s, K$
and the IRAC 3.6$\rm{\mu m}$, 4.5$\rm{\mu m}$, 5.8$\rm{\mu m}$ and
8.0$\rm{\mu m}$ bands; in the second panel, the SDSS $u, g, r, i, z$
bands; in the third panel, bands from HST instruments, three UV bands
from the WFC3-UVIS (0.225$\rm{\mu m}$, 0.275$\rm{\mu m}$, 0.336$\rm{\mu
  m}$), seven optical bands from the ACS-WFC (0.435$\rm{\mu m}$,
0.475$\rm{\mu m}$, 0.606$\rm{\mu m}$, 0.625$\rm{\mu m}$, 0.775$\rm{\mu
  m}$, 0.814$\rm{\mu m}$, 0.850$\rm{\mu m}$) and three near-infrared
bands from the WFC3-IR (1.05$\rm{\mu m}$, 1.25$\rm{\mu m}$,
1.60$\rm{\mu m}$); and in the bottom panel, the VIMOS U band, the 2
NICMOS near-infrared bands (1.1$\rm{\mu m}$ and 1.6$\rm{\mu m}$) and
two WFPC2 bands (0.30$\rm{\mu m}$ and 0.45$\rm{\mu m}$). While we will
not give results for all these bands in this paper, we will include
the relevant apparent magnitudes in our light-cone catalogues in order
to enhance their utility to others.

All magnitudes are in the AB system. In order to be as close as possible to
observations, we use the $K_s$-band when presenting results for number counts
and redshift distributions and the $K$-band when discussing the evolution of
the rest-frame luminosity function.

The lightcones constructed and made public in this work provide a
useful tool to test observational derivations of intrinsic galaxy
properties.  The wide wavelength coverage of observed- and rest-frame
photometry, together with the two stellar population synthesis models
considered, can be used to check derivations of rest-frame magnitudes
from observed photometry, as well as the reliability of properties
obtained from SED fitting, such as stellar masses, ages and
star-formation histories.

\section{Results}
\label{sec:results}

In this section we compare predictions of our models to observational
data. We start with number counts as a function of apparent magnitude
in a wide range of photometric bands (from the optical blue to the
IRAC bands) and move on to redshift distributions for $K$ and IRAC
$5.8\rm{\mu m}$ selected galaxies. Finally, we investigate the
evolution of the rest-frame optical and near-infrared luminosity
functions and colours which, although further from the directly
observed quantities, allow a better understanding of galaxy evolution.

\citet{Kitzbichler2007} and \citet{Torre2011} presented similar tests
for an earlier version of the Munich semi-analytic model. Here, we
take advantage of recent advances in the available observations and
extend these comparisons to higher redshift and to a wider range of
wavelengths. We also test the impact of population synthesis models by
comparing results for the \citet{Bruzual2003} and the
\citet{Maraston2005} models.  This follows up work by
\citet{Tonini2009, Tonini2010, Fontanot2010} and
\citet{Henriques2011}, who showed that the inclusion of near-infrared
emission from TP-AGB stars increases the predicted number of massive
and extremely red objects at $z\sim 2$, as seems to be required by
observation. Our comparison is based on a large number of lightcone
realizations with areas and selection effects matching the relevant
observational surveys.

\subsection{Number Counts}
\label{sec:nc}

\begin{figure*}
\centering
\includegraphics[width=17.9cm]{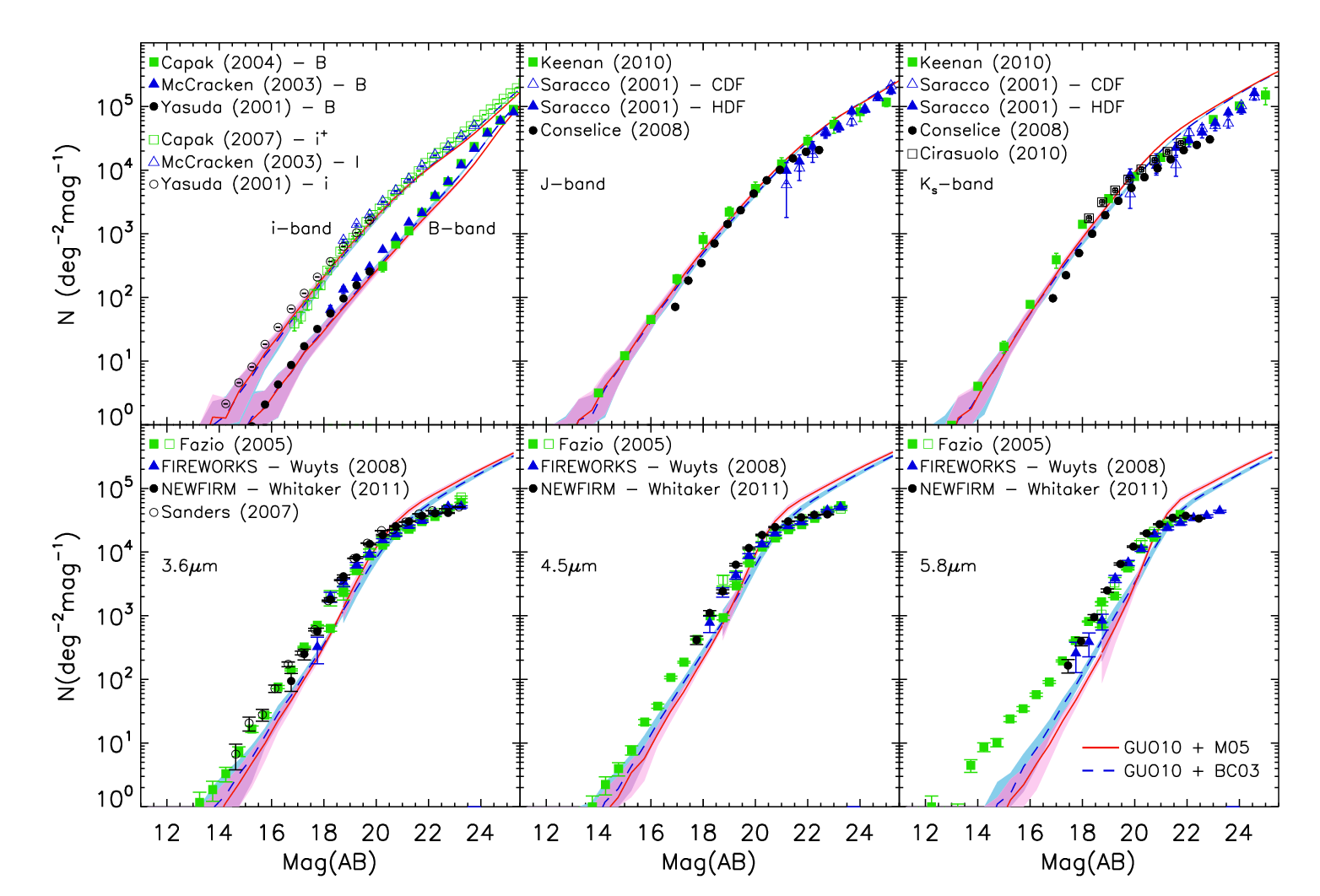}
\caption{Galaxy number counts as a function of apparent magnitude.  From top
  left to the bottom right, the panels show number counts in the $B$ and $i$,
  $J$, $K_s$, IRAC $3.6\rm{\mu m}$, $4.5\rm{\mu m}$ and $5.8\rm{\mu m}$
  bands. Theoretical predictions for the \citet{Maraston2005} and
  \citet{Bruzual2003} population synthesis models are shown as solid red and
  dashed blue lines respectively. The filled regions represent the 1$\sigma$
  field-to-field scatter for surveys of area 2~deg$^2$, except for the IRAC
  bands at faint magnitudes (M$>$18.5) where 100 arcmin$^2$ fields are
  assumed. The $B$-band number counts are compared with data from SDSS
  \citep{Yasuda2001}, VVDS \citep{McCracken2003} and HDF-N
  \citep{Capak2004}; $i$-band counts are also compared with data from SDSS and
  VVDS and with the COSMOS sample \citep{Capak2007};
  for the $J$- and $K_s$- bands we show observations from the CDF and HDF-S
  \citep{Saracco2001}, DEEP2 and Palomar \citep{Conselice2008} and MOIRCS
  \citep{Keenan2010} with UKIDSS-UDF data \citep{Cirasuolo2010} plotted 
  for the $K$-band; Spitzer \citep{Fazio2004}, FIREWORKS \citep{Wuyts2008} and
  NEWFIRM \citep{Whitaker2011} data are shown for the IRAC$_{3.6\rm{\mu m}}$,
  IRAC$_{4.5\rm{\mu m}}$ and IRAC$_{5.8\rm{\mu m}}$ bands.}
\label{fig:nc}
\end{figure*}

Galaxy counts in a given observed band can be difficult to interpret.
At each apparent magnitude they consist of galaxies at a wide range of
redshifts and thus with correspondingly wide ranges of absolute
magnitude and of emitted wavelength.  Nevertheless, such counts
provide an important test of models because they are directly observed
and so are independent of uncertainties in redshift, k-correction,
obscuration correction, etc.

Fig. \ref{fig:nc} shows galaxy number counts for the $B$ and $i$, $J$,
$K_s$, IRAC $3.6\rm{\mu m}$, $4.5\rm{\mu m}$ and $5.8\rm{\mu m}$ bands
(from top left to bottom right). Solid red and dashed blue lines are
model predictions for the \citet{Maraston2005} and \citet{Bruzual2003}
stellar population synthesis models respectively.  Filled regions show
the 1$\sigma$ field-to-field scatter expected among surveys of area
2~deg$^2$, except that 100 arcmin$^2$ fields are assumed for the IRAC
bands at faint magnitudes (M$>$18.5).

The optical $B$- and $i$-band number counts are compared with data
from the Sloan Digital Sky Survey \citep[SSDS;][]{Yasuda2001} and the
Vimos Very Deep Survey \citep[VVDS;][]{McCracken2003} at brighter
magnitudes. At fainter magnitudes we use the Hubble Deep Field - North
\citep[HDF-N;][]{Capak2004} for the $B$-band and the COSMOS sample
\citep{Capak2007} for the $i$-band. Both population synthesis models
match the data for bright galaxies (seen at low redshift in the
rest-frame optical), while the \citet{Bruzual2003} model predicts more
galaxies at faint apparent magnitudes, in better agreement with
observations. Similar trends were found for previous versions of the
Munich semi-analytic model \citep{Kitzbichler2007, Torre2011}. It is
difficult to draw firm conclusions from this result, however, since
these faint counts correspond to rest-frame ultra-violet emission from
galaxies at high redshift where uncertainties affect not only the
stellar population modelling, but also the simplistic treatments of
starbursts and of dust obscuration in the semi-analytic model.
\citet{Conroy2010} showed, for example, that increasing the number of
blue stragglers or blue horizontal branch stars increases the
predicted ultra-violet emission from passive galaxies. 

We compare the $J$- and $K_s$-band counts to observations from the
Chandra Deep Field and the Hubble Deep Field - South \citep[CDF and
HDF-S;][]{Saracco2001} from the DEEP2 and Palomar surveys
\citep{Conselice2008}, and from the MOIRCS sample \citep{Keenan2010}
. In addition, we show K-band counts based on the UKIDSS-Ultra Deep
Field data \citep[UKIDSS-UDF;][]{Cirasuolo2010}.  The two stellar
population synthesis models give similar predictions for the $J$-band
number counts which agree with the data. Both predict too many faint
objects in the $K_s$ and K bands. As shown by \citet{Guo2011}, and as
this paper will clarify, this is because the semi-analytic model
overpredicts the abundance of low-mass galaxies at high redshift.  The
two stellar population synthesis models predict similar counts at both
bright (low redshift, near infrared emission) and faint (high
redshift, red optical emission) apparent magnitudes, but they disagree
at intermediate apparent magnitudes. As we will see in more detail
below, the difference is a consequence of TB-AGB emission from stars
with ages of one or two Gyr which is fully included in the
\citet{Maraston2005} but not in the \citet{Bruzual2003} stellar
population model.

Predicted number counts for the IRAC $3.6\rm{\mu m}$, $4.5\rm{\mu m}$
and $5.8\rm{\mu m}$ bands are plotted against Spitzer
\citep{Fazio2004}, FIREWORKS \citep{Wuyts2008} and NEWFIRM
\citep{Whitaker2011} observations. Both the models and the
observations show a pronounced change in slope at an apparent
magnitude near 20, but the break is stronger in the observations than
in the model and occurs at slightly brighter apparent magnitudes. As a
result, the models under-predict the number of bright objects (low
redshift, emission longwards of the rest-frame K-band) and
over-predict the number of faint objects (high redshift, emission in
the rest-frame $JHK$ region). The latter underprediction is even more
pronounced here than in the K-band and again is likely due to the
overabundance of lower mass galaxies at $z\geq 1$ in the model. The
deficit of bright galaxies is visible also in the $z=0$ rest-frame
$K$-band luminosity function (Fig. \ref{fig:LfK}). Since
\citet{Guo2011} tuned their semi-analytic model to match the observed
low-redshift stellar mass function, this deficit implies overly large
mass-to-near-infrared-light ratios which might be explained by overly
small stellar metallicities. Indeed, \citet{Henriques2010} showed that
the most massive low-redshift galaxies in the \citet{DeLucia2007}
version of the model have stellar metallicities which are too low by
about a factor of two (the dashed red lines in their Figs 4 and 10).
An increase in metallicity could remove the discrepancy by reducing
the mass-to-near-infrared-light ratios in model.\footnote{The
  luminosity correction depends on the actual deficit in metallicity,
  which in turn depends strongly on which stellar population model is
  used to derive masses and metallicities for the observed galaxies.}
Another important factor, particularly for the IRAC $5.8\rm{\mu m}$
band, is the possible contamination by emission from hot dust,
specifically, emission in the 3.3, 6.2 and 7.7$\rm{\mu m}$ PAH
features \citep{Draine2007a, Draine2007b, DaCunha2008}. Such emission
is not included in the models but may well be significant in the real
low-redshift galaxies.

% This
%source of emission was found necessary to explain both the full SED
%from galaxies \citet{Cunha2008} and the correlation between
%near-infrared and optical resolved colors of nearby galaxies
%\citet{Zibetti2011}. Particularly the first IRAC bands can be affected
%by re-emission from hot dust and the 3.3, 6.2 and 7.7 $\rm{\mu m}$ PAH
%features \citep{Draine2007a, Draine2007b}.

%However, we note that
%\citet{Bruzual2003} predicts higher bright number counts when compared
%to \citet{Maraston2005}. This could indicate that the disagreement
%between model and data is in part due to uncertainties in the
%modelling of rest-frame IRAC emission from stellar population models.

\begin{figure*}
\centering
\includegraphics[width=17.9cm]{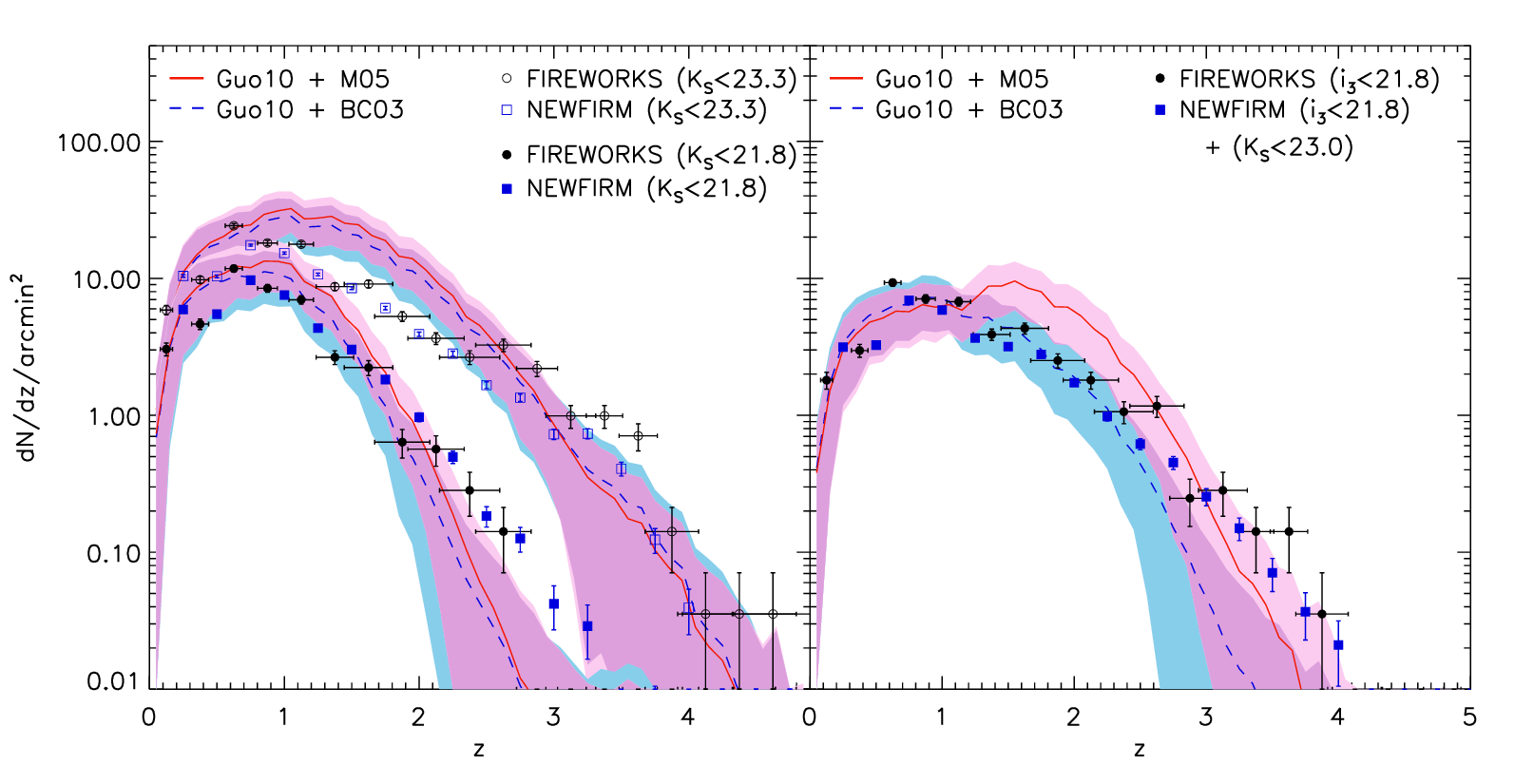}
\caption{The redshift distribution of galaxies selected above observed-frame
  $K_s$-band (left panel) and IRAC $5.8\rm{\mu m}$-band (right panel) apparent
  magnitude limits.  The solid red and dashed blue lines represent the mean
  predictions of our semi-analytic model for the \citet{Maraston2005} and
  \citet{Bruzual2003} stellar population synthesis models. The filled regions
  show the 1$\sigma$ scatter among fields with area 100 arcmin$^2$. The left
  panel shows the total number of galaxies per unit area and redshift for
  samples with $K_s<21.8$ and $K_s<23.3$. Similar curves are shown in the
  right panel but for samples with IRAC$_{5.8\rm{\mu m}}<$21.8 and
  $K_s<$23.0. Theoretical distributions are compared with data from FIREWORKS
  \citep{Wuyts2008} and from the NEWFIRM Medium Band
  Survey \citep{Whitaker2011}.}
\label{fig:zdist}
\end{figure*}

\subsection{Redshift distributions for $K$ and IRAC$_{5.8\rm{\mu m}}$ selected samples}
\label{sec:zdist}

In Fig. \ref{fig:zdist} we compare predictions from our models to the
observed photometric redshift distributions of galaxy samples
selected above $K_s$ and IRAC $5.8\rm{\mu m}$ apparent magnitude
limits.  As for the number counts, the solid red and dashed blue lines
represent predictions based on the \citet{Maraston2005} and
\citet{Bruzual2003} stellar population synthesis models. Filled
regions show the expected 1$\sigma$ scatter among fields with an area
of 100 arcmin$^2$. In the left panel the number of galaxies per unit
area and redshift is plotted for samples with $K_s<21.8$ and
$K_s<23.3$, while the right panel gives similar results but for samples
with IRAC $5.8\rm{\mu m}$ apparent magnitude brighter than 21.8 (and
$K_s<23.0$). Our theoretical predictions are compared with data from
two public catalogues: the FIREWORKS data for the GOODS-CDF
\citep{Wuyts2008} and the NEWFIRM Medium Band Survey data for the
COSMOS and AEGIS fields \citep{Whitaker2011}. The wide photometric
coverage of these two datasets results in robust and relatively
precise photometric redshift measurements. Our theoretical samples are
selected using photometric criteria very similar to those defining the
observed samples, although we plot the distribution of their true
redshifts rather than attempting to reproduce the observational
redshift estimation procedure.

Selection by observed-frame $K_s$-band magnitude picks galaxies on the
basis of their rest-frame $K$-band emission at low redshift, their
rest-frame $J$ band emission at $z\sim 1$, and their rest-frame
optical emission at $z>1.5$. For both magnitude limits, the samples
are dominated by intrinsically faint objects at low-redshift but by
intrinsically bright galaxies beyond $z\sim1$ (for $K_s<21.8$) or
$z\sim1.5$ (for $K_s<23.3)$).  For both apparent magnitude limits the
predictions for the two population synthesis models agree, and for the
brighter limit they are consistent with the observations out to a
redshift of almost two. Both underpredict the counts at higher
redshift, with the effect being slightly larger for the
\citet{Bruzual2003} model. This may reflect an underabundance of
intrinsically bright objects (in the optical) in the model at these
redshifts, but could also be due to magnitude and photometric redshift
errors in the data which primarily affect the tails of the
distribution.  For the fainter apparent magnitude limit the model
clearly overestimates the number of objects over the redshift range
$1<z<2.5$. These galaxies typically have stellar masses of order a few
$10^{10}M_\odot$ and this discrepancy reflects the overabundance of
objects of this mass and redshift unity that was flagged by
\citet{Guo2011}. The same problem was identified in earlier versions
of the model by \citet{Kitzbichler2007} and \citet{Torre2011}. At
higher and lower redshifts the abundances agree quite well in model
and data, reflecting the fact that the semi-analytic model was tuned
to fit galaxy abundances at low redshift, and predicts an abundance of
high-mass galaxies which fits observed estimates quite well at high
redshift.

For the IRAC $5.8\rm{\mu m}$ selected samples shown in the right panel
of Fig.  \ref{fig:zdist}, there is a significant difference between
the predictions of the two population synthesis models. While the
\citet{Bruzual2003} model predicts a distribution with similar shape
to those in the left panel, the \citet{Maraston2005} model makes a
concordant prediction only at $z<1.2$.  Beyond this point there is a
``bump'' and at higher redshift it predicts roughly 3 times as many
galaxies as the \citet{Bruzual2003} model. A corresponding bump is not
present in the observational data which are better described by the
\citet{Bruzual2003} model, at least out to $z\sim 2.5$. The bump in
the \citet{Maraston2005} model is caused by strong rest-frame $JHK$
emission from TP-AGB stars associated with intermediate age stellar
populations. While this emission brings the predicted numbers of
galaxies into rough agreement with the data at the highest redshifts,
it results in an overabundance at $z\sim 2$ where the observational
datasets appear most robust. Since this effect is also present for
data in other wavebands, for which TP-AGB emission is not an issue, it
suggests that it results from the semi-analytic model overpredicting
the abundance of the relevant moderate mass galaxies by a substantial
factor at this redshift. In view of this, the agreement achieved by
the \citet{Bruzual2003} model is probably coincidental, resulting from
the overestimated abundance of moderate mass galaxies being
compensated by an overestimate of their rest-frame $JHK$ mass-to-light
ratios.

\begin{figure*}
\centering
\includegraphics[width=17.9cm]{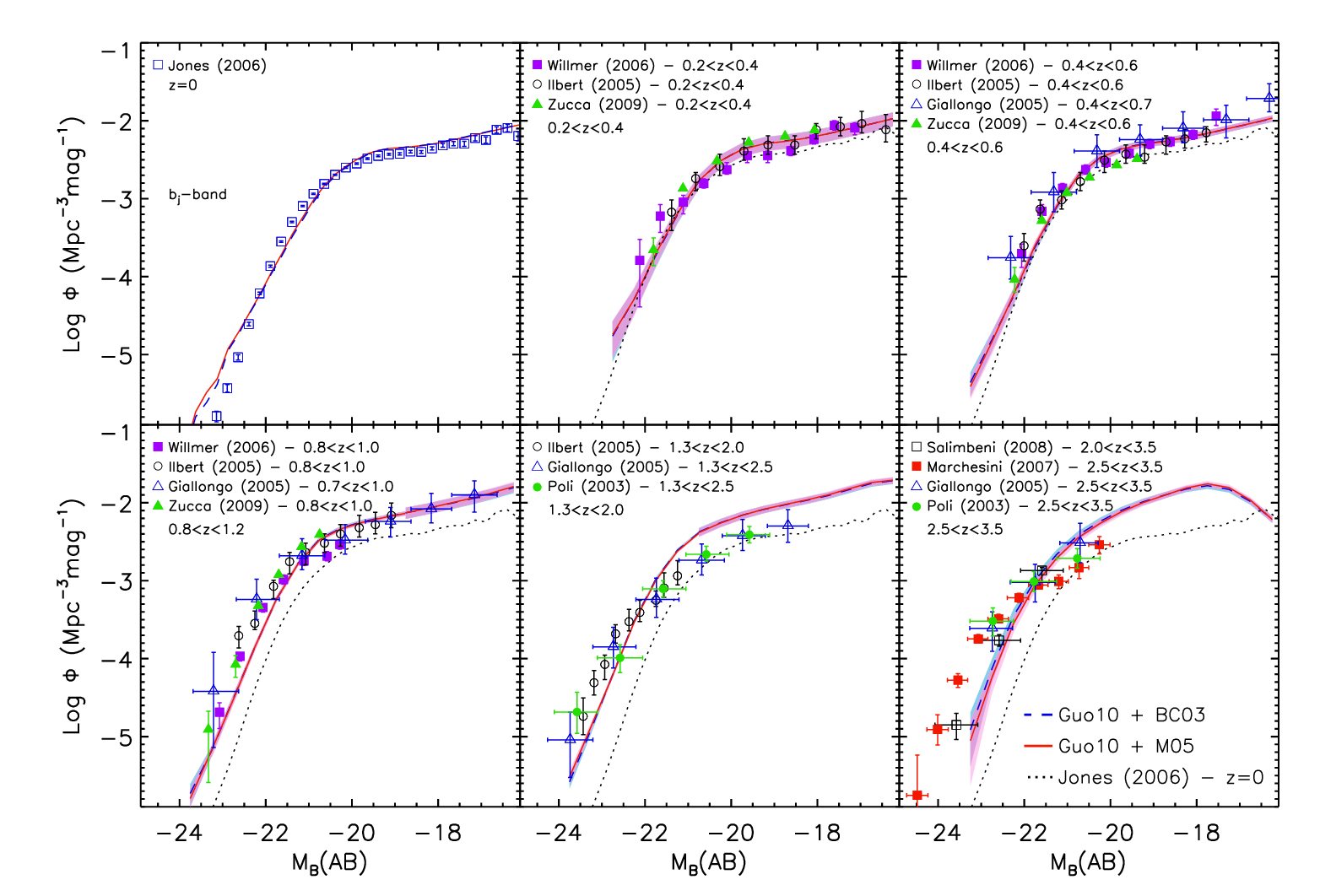}
\caption{Evolution of the rest-frame $B$-band luminosity function from
  $z=3$ to $z=0$. Theoretical predictions for the \citet{Maraston2005}
  and \citet{Bruzual2003} stellar population models are shown as solid
  red and dashed blue lines respectively. Filled regions represent the
  1$\sigma$ field-to-field scatter expected for surveys of area 1.4
  deg$^2$, except in the highest redshift panel, where 150 arcmin$^2$
  fields are assumed. Fields of this size are also assumed for the
  intrinsically fainter galaxies in the 0.8$<$z$<$1.2 and
  1.3$<$z$<$2.0 panels (for galaxies with M$_B>-19.0$ and $>-21.0$
  respectively). At $z=0$ the model $b_j$-band luminosity function
  \citep[$b_j=B-0.267(B-V)$;][]{Norberg2002} is compared with
  observations from the 6DFGRS \citep{Jones2006}, repeated at all
  redshifts as a dotted black line. At higher redshifts we show
  observational estimates from VVDS \citep{Ilbert2005}, DEEP 2
  \citep{Willmer2006}, zCOSMOS \citep{Zucca2009}, HDF-S
  \citep{Poli2003}, HDF-N \citep{Giallongo2005}, GOODS-MUSYC plus
  FIRES \citep{Marchesini2007} and GOODS-MUSYC \citep{Salimbeni2008}.}
\label{fig:LfB}
\end{figure*}

\subsection{The rest-frame $B$-Band Luminosity Function}
\label{sec:LfB*}

Rest-frame luminosity functions and colour distributions as a function of
redshift provide direct estimates of the abundance evolution of various galaxy
types (e.g. star-forming/passive, high/low mass). However, they require
accurate redshifts and appropriate photometry if they are to be determined
reliably from observed-frame fluxes. The wide wavelength coverage of modern
surveys produces robust photometric redshifts, and, in addition, allows
rest-frame optical and near-infrared magnitudes to be determined by
interpolation over the full range $0<z<4$, rather than requiring an uncertain
extrapolation based on an SED fit. 

\citet{Guo2011} showed that their semi-analytic model reproduces
observed $z\sim 0.1$ luminosity functions in the SDSS $g, r, i$ and
$z$ bands. At higher redshift they implemented the redshift-dependent
dust model of \citet{Kitzbichler2007}. This reproduces the observed
abundance of colour-selected galaxies at $z\sim 2$ and $z\sim 3$ for
the previous version of the semi-analytic model
\citep[see][]{Guo2009}. In Fig. \ref{fig:LfB} we show the evolution of
the $B$-band luminosity function from $z=0$ to $z=3$ for our current
semi-analytic model. Solid red and dashed blue lines, represent
versions with the \citet{Maraston2005} and \citet{Bruzual2003} stellar
population models respectively.  Filled regions give the expected
1$\sigma$ field-to-field scatter for surveys of area 1.4~deg$^2$,
except that smaller fields (with area 150~arcmin$^2$) were assumed for
the $2.5<z<3.5$ panel, for the $1.3<z<2.0$ panel fainter than
$-21.0$ and for the $0.8<z<1.2$ panel fainter than $-19.0$.

At $z\sim 0$ the model $b_j$-band luminosity function\footnote{We
  assume $b_j=B-0.267(B-V)$, \citep{Norberg2002}.} is compared with the
6 Degree Field Galaxy Redshift Survey result of
\citet[][6DFGRS]{Jones2006}, repeated for reference as a black dotted
line in all panels. For the $z\sim 0$ panel we use the final snapshot
of the simulation rather than the lightcone, finding excellent
agreement with the 6DFRS result, just as was the case for the
corresponding SDSS luminosity function (in the $g$ band) in
\citet{Guo2011}. For the other $z\leq 1$ panels we compare with
data from the relatively wide VVDS \citep{Ilbert2005}, DEEP2
\citep{Willmer2006} and zCOSMOS \citep{Zucca2009} surveys. At higher
redshift only data for smaller fields are available.  We show results
from \citet[][HDF-S]{Poli2003}, \citet[][HDF-N]{Giallongo2005},
\citet[][GOODS-MUSYC+FIRES]{Marchesini2007} and
\citet[][GOODS-MUSYC]{Salimbeni2008}.

The model reproduces the evolution of the rest-frame $B$-band
luminosity function reasonably well out to $z=3$. It overpredicts the
abundance of faint objects at $z\sim2$ and underpredicts the abundance
of bright objects at $z\sim3$, although it may still be compatible
with the data given the relatively large error bars quoted by the
observers and the substantial scatter between the observational
determinations. The two stellar population synthesis models give very
similar results in this band. We note that the predicted fluxes are
strongly affected by dust, and so are dependent on the adopted dust
model. Further testing of the simplistic and relatively poorly
motivated model of \citet{Kitzbichler2007} is clearly needed.

\begin{figure*}
\centering
\includegraphics[width=17.9cm]{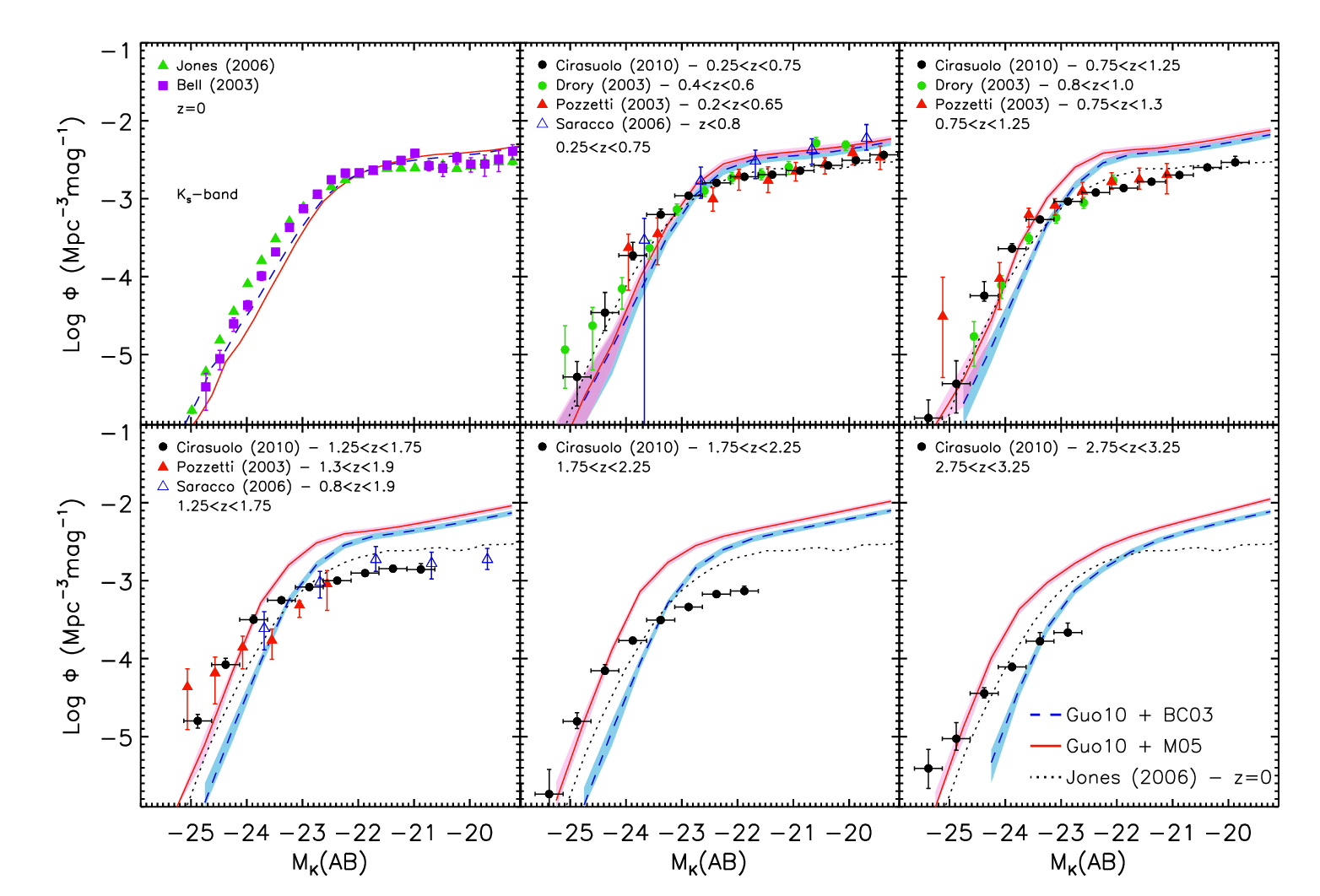}
\caption{Evolution of the rest-frame $K$-band luminosity function from $z=3$
  to $z=0$. Predictions of our semi-analytic model for the
  \citet{Maraston2005} and \citet{Bruzual2003} population synthesis models are
  shown by solid red and dashed blue lines respectively. Filled regions
  represent the expected 1$\sigma$ field-to-field scatter for surveys of area
  0.7~deg$^2$. At $z\sim0$ the model $K_s$-band luminosity function is
  compared with data from 2MASS \citep{Bell2003} and 6DFGRS+2MASS
  \citep{Jones2006}. We repeat the latter at all redshifts as a black
  dotted line. At higher redshifts we show observational estimates based on
  MUNICS \citep{Drory2003}, the UKIDSS-UDF \citep{Cirasuolo2010}, the K20
  Survey \citep{Pozzetti2003} and the HDF-S \citep{Saracco2006}. Note that in
  all surveys other than the UKIDSS-UDF, the rest-frame K-luminosities are not
  directly measured but are rather estimated by extrapolating from the {\it
    observed frame} $K$-band fluxes using an SED model.}
\label{fig:LfK}
\end{figure*}

\subsection{The $K$-Band Luminosity Function}
\label{sec:LfK}

The $K$-band luminosity function has long been thought of as a proxy for the
stellar mass function. Recent results have shown, however, that this
assumption, while moderately accurate at low redshift, can break down badly at
early times. Notably, the fact that the characteristic luminosity $L_\star$
increases with increasing redshift just as for the optical bands
\citep{Cirasuolo2010} is inconsistent with a time-independent $K$-band
mass-to-light ratio, which would imply the ``evaporation'' of material from
the most massive galaxies. This luminosity function behaviour is easily
understood in the context of recent stellar population synthesis models. In
particular, a significant amount of $K$-band emission comes not from the old
populations which dominate the stellar mass, but rather from intermediate age
stars ($\sim$1 Gyr) passing through the TB-AGB phase \citep{Maraston2005,
  Charlot2007}.  At high redshifts these relatively young populations can
dominate the rest-frame luminosity in the $K$-band and they are only later
replaced by predominantly old populations \citep{Henriques2011}.

In Fig. \ref{fig:LfK} we plot the evolution of the $K$-band luminosity
function out to $z=3$. As in previous figures, the solid red and
dashed blue lines represent predictions based on the
\citet{Maraston2005} and \citet{Bruzual2003} stellar population
models, and filled regions outline the expected 1$\sigma$
field-to-field scatter among surveys of area 0.7~deg$^2$. At $z\sim0$
the model $K_s$-band luminosity function is compared with
observational data from 2MASS \citep{Bell2003} and 6DFGRS+2MASS
\citep{Jones2006}. As a reference, we repeat the latter at all
redshifts as a black dotted line. For the $z\sim0$ panel we use the
final snapshot of the simulation rather than the lightcone to obtain
the theoretical prediction.  At higher redshifts, the model is
compared with data from the MUNICS \citep{Drory2003}, K20
\citep{Pozzetti2003}, HDF-S \citep{Saracco2006} and UKIDSS-UDF
\citet{Cirasuolo2010}. Only the last of these surveys estimates the
rest-frame $K$-band flux directly by interpolating between
observed-frame magnitudes at corresponding wavelengths (from
Spitzer/IRAC). The other surveys extrapolate the {\it observed-frame}
$K$ flux to longer wavelength using an uncertain SED fit and thus may
be subject to substantial systematic errors. 

\begin{figure*}
\centering
\includegraphics[width=17.9cm]{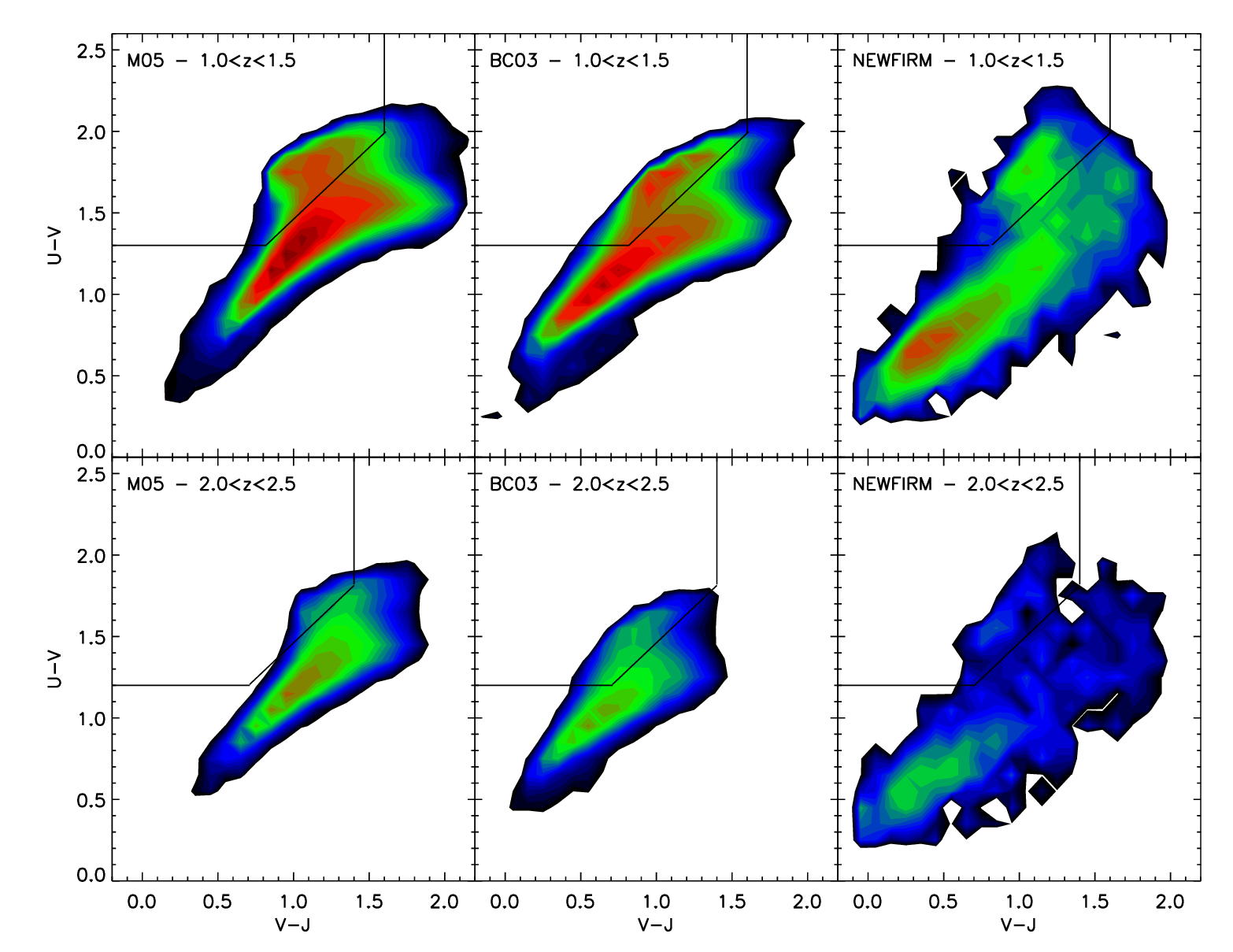}
\caption{The rest-frame $U$-$V$ versus $V$-$J$ colour diagram for
  $1.0<z<1.5$ (top panel) and $2.0<z<2.5$ (bottom panel). Model
  predictions for the \citet{Maraston2005} and \citet{Bruzual2003}
  stellar populations (respectively left and middle panels) are
  compared with NEWFIRM data from \citet{Whitaker2011} (right
  panel). Theoretical galaxies were selected to have observed $K<23.0$
  in order to match the observational selection. The contours
  represent the density of points with the total number of objects
  normalized by the area surveyed. The solid black
  line shows the observational dividing line between active and passive
  objects.}
\label{fig:colours}
\end{figure*}

Our two population synthesis models give very similar predictions for the
rest-frame $K$-band luminosity function out to $z\sim0.5$. At higher redshifts
their shapes and their normalisations remain similar but their characteristic
luminosities diverge with the \citet{Maraston2005} prediction being brighter
by about 0.4, 0.5, 0.6 and 0.75 magnitudes at redshifts of 1.0, 1.5, 2.0 and
3.0 respectively. This reflects the increasing contribution from TP-AGB stars
as the mean age of the galaxies gets younger. The predictions of both models
are strongly at variance with observation at these redshifts. While the
\citet{Maraston2005} model agrees with the high-mass tail of the observed
luminosity functions at all redshifts, it seriously overpredicts the abundance
of less massive galaxies at $z= 1$ and earlier. For the lowest luminosity bin
of the UKIDSS-UDF dataset the overprediction is by factors of 2, 4, 6 and
almost 8 at redshifts of 1.0, 1.5, 2.0 and 3.0 respectively. The
\citet{Bruzual2003} model fails to reproduce the rest-frame $K$ luminosities
of the most massive systems (by about 0.7 mag by $z\sim3$) but nevertheless
overpredicts the abundance of less luminous systems almost as badly as the
\citet{Maraston2005} model. This is the substantial problem already pointed
out by \citet{Guo2011}; their galaxy formation assumptions produce moderate
mass galaxies ($M_\star\sim 10^{10}M_\odot$) too early to be compatible with
current data on populations at $z\geq 1$.

As noted above when discussing Fig. \ref{fig:nc}, the model also slightly
underestimates the $K$-band luminosities of massive low-redshift
galaxies. This more subtle problem is due to these massive galaxies being too
blue (see the colour distributions in Fig.~12 of \citealt{Guo2011}). This is
likely caused by an underabundance of heavy elements \citep{Henriques2010}.

Similar results for the evolution of the rest-frame $K$-band
luminosity function have been obtained for an earlier version of the
Munich semi-analytic model (\citealt{DeLucia2007}, as well as for the
semi-analytic models of \citealt{Menci2006}, \citealt{Monaco2007} and
\citealt{Fontanot2009}). In a recent paper \citet{Somerville2011}
compared another independent semi-analytic model to observational data
on the evolution of the rest-frame $1500\AA$, $B$ and $K$ band
luminosity functions. The authors were able to get a reasonable match
to the bright tail without including TP-AGB emission, but they do
obtain a much weaker evolution of the characteristic $L_{\star}$ than
observed, in concordance with our \citet{Bruzual2003} results. They
also overpredict the abundance of lower luminosity galaxies by very
similar factors to those that we find here. It seems that whatever is
causing the overly early formation of lower mass galaxies is common to
all recent semi-analytic models.

\subsection{Galaxy Colours}
\label{sec:LfK}

Recent observations have shown that the local bimodality between blue,
star-forming and red, passive galaxies persists at least up to $z=2$
(e.g. \citealt{Wuyts2007}; \citealt{Williams2009};
\citealt{Ilbert2010}; \citealt{Whitaker2011}). These authors have used
a combination of a rest-frame near-infrared colour and an optical
color in order to separate dusty star-forming galaxies from passive
objects.  At fixed $U$-$V$, red passive galaxies will have bluer
$V$-$J$ colors than dusty star-forming objects.  In
Fig. \ref{fig:colours} we plot rest-frame $U$-$V$ versus $V$-$J$
diagrams in two redshift bins, $1.0<z<1.5$ in the upper panels and
$2.0<z<2.5$ in the bottom panels. Predictions are shown for two
different stellar populations (\citet{Maraston2005} in the left panels
and \citet{Bruzual2003} in the middle panels) and for NEWFIRM
observations \citep{Whitaker2011}. Theoretical galaxies in the
lightcones were selected to match observations by applying a flux
limit at observed $K$-band $=23.0$, roughly the 90\% completion limit
quoted for observations. As described in \citet{Whitaker2011},
observational galaxies were carefully deblended and only objects with
S/N$>$8 in the $K$ band were included. The contours represent the
density of points with the total number of objects normalized by the
area surveyed. The solid black line shows the empirical dividing line
between active and passive objects.

For both redshift bins, the models correctly predict the existence of
two distinct populations, although they fail to match the exact
observational spread in colour. The two populations have less scatter
and are closer to each other in the models also covering a smaller
range in $V$-$J$. This might in part result from incorrect physics in
the model but it can also be explained by uncertainties in the
conversion between aperture and total magnitudes, photometric
redshifts and the process of SED fitting when deriving total
rest-frame magnitudes from observations. At $z=2$, the colours of red
galaxies seem to be better matched by the \citet{Bruzual2003}
prescription. The \citet{Maraston2005} predictions are shifted to
larger $V$-$J$ colours. Nevertheless, we note that the position of a
galaxy population in this diagram is strongly dependent on the dust
model assumed. For both population models, passive galaxies do not
form a distinct peak, but rather a cloud of objects departing from the
blue sequence towards redder $U$-$V$ colours (at $1.0<V-J<1.5$ for
\citet{Maraston2005} and $0.5<V-J<1.0$ for \citet{Bruzual2003}). These
are in fact passive galaxies in the model with almost no on-going star
formation.

The current model for galaxy formation in a hierarchical Universe
predicts the most massive galaxies to grow rapidly at the centres of
clusters and large groups. Their history is rich in merger events
which can fuel gas into their central black holes. Feedback from these
objects can then shut down star formation at early times. The redder
objects that can be seen at $z=2$ in the model have masses between
$10^{11}\Msun$ and $10^{11.5}\Msun$ and black hole masses as big as
$10^{8}\Msun$.

\section{Conclusions}
\label{sec:conclusions}

We have constructed lightcones from the latest version of the Munich
semi-analytic model \citep{Guo2011} and used them to compare the model
with the high-redshift galaxy population as revealed by recent deep
surveys at optical and near-infrared wavelengths. We have combined the
model with two different stellar population synthesis packages
\citep{Bruzual2003, Maraston2005} in order to understand how
differences in the photometric modelling are reflected in inferences
about galaxy evolution.  We use multiple independent lightcones to
characterize cosmic variance uncertainties in currently available
datasets.  Our mock catalogues are made publicly available and provide
observer-frame photometry in 40 commonly used photometric bands, in
addition to rest-frame photometry and a variety of physical properties
of the galaxies (positions, peculiar velocities, stellar masses, halo
masses, sizes, morphologies, gas fractions, star formation rates,
metallicities, halo properties).

We now summarise the principal conclusions from our comparison of models and
data.
\begin{itemize}
\item For both stellar populations the model matches the observed-frame $B, i$
  and $J$ number counts but overpredicts the counts at faint magnitudes
  ($m_{AB}>20$) in the $K_s$, IRAC $3.6\rm{\mu m}$, $4.5\rm{\mu m}$ and
  $5.8\rm{\mu m}$ bands. This reflects the overproduction of moderate mass
  galaxies (stellar masses $M_\star\sim10^{10}M_\odot$) at $z\geq 1$ already
  noted by \citet{Guo2011}. The matching of the faint optical counts is
  fortuitous -- this overproduction is masked by overly large mass-to-light
  ratios in the rest-frame near-UV, perhaps due to problems with the dust
  modelling.

\item At bright magnitudes ($m_{AB}<20$) the model underpredicts the counts in
  the three IRAC bands. This is due to an underestimation of the near-infrared
  luminosities of low-redshift massive galaxies caused in part by the fact
  that such galaxies are insufficiently metal-rich in the model, and in part
  by the model's neglect of PAH emission from hot dust (which is particularly
  significant at $5.8\rm{\mu m}$).

%For the IRAC bands,
 % there is a deficit of bright objects which we interpret as
 % contamination from dust re-emission in observations as well as
 % uncertainties in stellar populations.

\item At magnitudes where the model $K_s$-band counts agree with 
  observations the redshift distribution of $K_s$-selected samples is also
  reproduced. At fainter magnitudes where the counts are overpredicted, the
  excess galaxies occur primarily at $1<z<2.5$, again reflecting the
  overproduction  of $M_\star\sim 10^{10}M_\odot$ galaxies at these epochs.
  The two population synthesis models give similar results in both regimes.

\item The two population synthesis models predict different redshift
  distributions for galaxies selected to $m_{AB}\sim 22$ in the IRAC
  $5.8\rm{\mu m}$ band.  Emission from TP-AGB stars enhances the number of
  galaxies at $z>1.5$ in the \citet{Maraston2005} model by about a factor of 3
  relative to the \citet{Bruzual2003} model, causing it to overpredict the
  observed abundance at $z\sim 2$. Although the \citet{Bruzual2003} model
  agrees with the observed redshift distribution for $0<z<3$, this is a result
  of the overabundance of moderate mass galaxies being cancelled by an
  overestimate of their near-infrared mass-to-light ratios.

\item The two population synthesis give similar results for the evolution of
  the rest-frame $B$-band luminosity function, agreeing well with observation
  out to $z=1.2$. At higher redshift the agreement is less convincing. The
  overabundance of lower mass model galaxies starts to become evident, and
  there is some indication that the models underpredict the abundance of the
  most luminous objects. Cosmic variance and other uncertainties in the
  currently available data, together with dust modelling uncertainties in the
  model, preclude any strong conclusions.

\item The \citet{Maraston2005} population model reproduces the bright tail of
  the rest-frame $K$-band luminosity function all the way out to $z\sim 3$,
  whereas the \citet{Bruzual2003} model underpredicts the near-infrared
  luminosities of these massive galaxies by an amount which increases from
  about 0.3~mag at $z\sim 1$ to 0.7~mag at $z\sim 3$. The overproduction of
  $M_\star\sim 10^{10}M_\odot$ galaxies at these times causes both models to
  substantially overpredict galaxy abundances below the knee of the luminosity
  function.

\item The model predicts that a population of red, passive
    galaxies galaxies should be in place already at $z=2$, as seen in
    observations. These are the most massive galaxies at the centres
    of clusters and large groups which can rapidly grow a central
    black hole capable of producing enough feedback to stop star
    formation at early times.

\end{itemize}

In the literature it has often been suggested that semi-analytic
models fail to reproduce the rest-frame $K$-band galaxy luminosities
of the brightest high-redshift galaxies (at $z\sim~2-3$), and the
failure is usually attributed to insufficiently rapid mass growth at
early times \citep{Pozzetti2003, Cimatti2004, Kitzbichler2007,
  Cirasuolo2010}. This study (and that of \citealt{Henriques2011})
suggest otherwise -- current models seem fully capable of reproducing
the data given realistic assessments of population synthesis
uncertainties and of the effects of observational errors. A much more
serious problem, as already pointed out in the literature
(e.g. \citealt{Marchesini2009, Marchesini2010}, \citealt{Guo2011},
\citealt{Somerville2011}), is that the models grow somewhat lower mass
galaxies too early. Objects with $M_{\star}\sim 10^{10}M_\odot$ are
already present with a large fraction of their $z=0$ abundance at
redshifts of 2 or 3, whereas the observations indicate a drop in
abundance by about an order of magnitude. Cosmic down-sizing thus
appears much stronger in the real Universe than in the
models. Reconciling theory and observation in the context of the
$\Lambda$CDM cosmology will require star formation efficiencies to
scale with mass and redshift in a very different way than current
models (and simulations) assume.

\section*{Acknowledgements}
The halo/subhalo merger trees for the Millennium and Millennium-II
Simulations as well as galaxy catalogues implemented on these
simulations are publicly available at
http://www.mpa-garching.mpg.de/millennium.  This interface was created
as part of the activities of the German Astrophysical Virtual
Observatory \citep{Lemson2006}. The light-cone mock catalogues
constructed for this work are also public at the same location. The
work of BH, SW and GL was supported by Advanced Grant 246797
``GALFORMOD'' from the European Research Council. PT was supported by
the Science and Technology Facilities Council [grant number
ST/I000976/1]. GQ acknowledges support from the National basic
research program of China (program 973 under grant No. 2009CB24901),
the Young Researcher Grant of National Astronomical Observatories,
CAS, the NSFC grants program (No. 11143005) and the Partner Group
program of the Max Planck Society, as well as the hospitality of the
Institute for Computational Cosmology in Durham, UK. The authors thank
Jeremy Blaizot for providing the MoMaF code, Stijn Wuyts and Katherine
Whitaker for the use of their observational data and the anonymous
referee for helpful comments.

\bibliographystyle{mn2e} \bibliography{paper}

\label{lastpage}

\end{document}